\documentclass{PoS}
\usepackage{amssymb}
\usepackage{bm}
\usepackage{latexsym}

\newcommand{\case}[2]{\ensuremath{{\textstyle\frac{#1}{#2}}}}
\newcommand{\slsh}[1]{\ensuremath{{#1\kern -0.55em /}}}

\newcommand{\Dirac}{\ensuremath{{D\kern -0.65em /}}}

\newcommand{\half}{\ensuremath{{\textstyle\frac{1}{2}}}}

\newcommand{\quarter}{\ensuremath{{\textstyle\frac{1}{4}}}}

\newcommand{\third}{\ensuremath{{\textstyle\frac{1}{3}}}}

\renewcommand{\Re}{\mathop{\mathrm{Re}}}

\title{Non-Standard Physics in Leptonic and Semileptonic Decays of 
Charmed Mesons}

\ShortTitle{Non-Standard $\bar{s}c\bar{\nu}\ell$}

\author{\speaker{Andreas S. Kronfeld} \\
		Theoretical Physics Department,
		Fermi National Accelerator Laboratory,$\!$%
		\thanks{Operated by Fermi Research Alliance, LLC, under Contract
		No.~DE-AC02-07CH11359 with the United States Department of Energy.}~
		Batavia, Illinois, USA \\
		E-mail: \email{ask@fnal.gov}}

\abstract{Recent measurements of the branching fraction for 
$D_s\to\ell\nu$ disagree with the Standard Model expectation, which 
relies on calculations of $f_{D_s}$ from lattice QCD.
This paper uses recent preliminary measurements from CLEO and a new 
preliminary lattice-QCD result from this conference to update the 
significance of the discrepancy.
The ``$f_{D_s}$ puzzle'' stands now at~3.5$\sigma$, with $\sigma$
predominantly from the statistical uncertainty of the experiments.
New physics scenarios that could solve the puzzle would also lead to
non-Standard amplitudes mediating the semileptonic decays
$D\to K\ell\nu$.
This paper shows where the new amplitudes enter the differential rate 
and outlines where lattice QCD calculations are needed to confront 
recent and forthcoming measurements.}

\FullConference{The XXVI International Symposium on Lattice Field Theory \\
	July 14 - 19, 2008  \\ Williamsburg, Virginia, USA}

\begin{document}

\section{Introduction}

Recent years have witnessed significant improvements in charmed-meson 
leptonic and semileptonic decays, both in experimental measurements and 
in calculations of the hadronic transition amplitudes with lattice QCD.
A~puzzle has arisen, namely a discrepancy of approximately 3.5$\sigma$
in the rate of the leptonic decay $D_s\to\ell\nu$, where $\ell$ is a
muon or $\tau$~lepton~\cite{Rosner:2008yu}.
If the measured counts have not fluctuated high, and the lattice QCD
calculations are confirmed (by further calculations with 2+1 flavors of
sea quarks), then this may be a signal of physics beyond the Standard
Model~\cite{Dobrescu:2008er}.

If non-Standard interactions mediate $c\bar{s}\to\nu\bar{\ell}$, then 
they also alter, at some level, the rate and $q^2$-distribution of 
$D\to K\mu\nu$.
($D\to K\tau\nu$ is kinematically forbidden.)
In this paper, section~\ref{sec:fds} recalls the origin of the leptonic
discrepancy, incorporating new, preliminary results.
Section~\ref{sec:np} updates the new-physics analysis of
Ref.~\cite{Dobrescu:2008er} and extends it to encompass semileptonic decays.
Then section~\ref{sec:semileptonic} discusses the phenomenology of
semileptonic decays in the context of new physics.
For lattice QCD the main conclusion, discussed in section~\ref{sec:end},
is that precise calculations of the semileptonic form factors, including
a tensor form factor defined below, are vital.

\section{Leptonic Decays}
\label{sec:fds}

In the Standard Model the partial width for $D_s\to\ell\nu_\ell$ is
\begin{equation}
	\Gamma(D_s\to\ell\nu_\ell) = \frac{m_{D_s}}{8\pi} f_{D_s}^2
		\left|G_FV_{cs}^*m_\ell\right|^2
		\left(1-m_\ell^2/m_{D_s}^2\right)^2 ,
	\label{eq:Dslnu}
\end{equation}
where the decay constant $f_{D_s}$ is defined by
$\langle0|\bar{s}\gamma^\mu\gamma_5c|D_s(p)\rangle=if_{D_s}p^\mu$, 
and is also computed via
$(m_c+m_s)\langle0|\bar{s}\gamma_5c|D_s(p)\rangle=-if_{D_s}m_{D_s}^2$;
PCAC ensures that the two definitions are the same.
The partial widths are small:
for muonic decays owing to the helicity-suppression factor $m_\mu^2$;
for $\tau$-leptonic decays owing to the phase-space factor 
$(1-m_\tau^2/m_{D_s}^2)^2$.
Experiments measure the branching fraction $B=\Gamma\tau_{D_s}$ but
usually quote $f_{D_s}$ assuming that no non-Standard amplitude
contributes to~$\Gamma$.

In this sense, $f_{D_s}$ has been measured recently by the
BaBar~\cite{Aubert:2006sd}, Belle~\cite{Abe:2007ws}, and
CLEO~\cite{Pedlar:2007za,Ecklund:2007zm} Collaborations.
The experiments measure $B(D_s\to\ell\nu)$ directly, without complicated
modeling of the events or background, and the experimental errors are
principally statistical.
Radiative corrections are at most 1--2\%, and the discrepancy cannot be
explained with any value of $|V_{cs}|$ consistent with a unitary
$n\times n$ CKM matrix~\cite{Dobrescu:2008er}.
In summary, it seems sound to take the experimental measurements of
$f_{D_s}$ at face value, yielding Table~\ref{tab:fDs}.
\begin{table}[tp]
	\centering
	\caption{Recent experimental values of $f_{D_s}$.
	The preliminary update from CLEO can be found in Ref.~\cite{CLEO:new}.}
	\label{tab:fDs}
	\begin{tabular}{l@{\quad}l@{\quad}l@{\quad}l}
		\hline\hline
		final state & reference & \multicolumn{2}{c}{$f_{D_s}$~(MeV)}  \\
		            &           & end 2007 & 2008 update \\
		\hline
		$\mu\nu$  & BaBar~\cite{Aubert:2006sd} & $283\pm17\pm16$ & \\
		$\mu\nu$  & CLEO~\cite{Pedlar:2007za}  & $264\pm15\pm\;7$ & 
		$265.4\pm11.9\pm4.4$ \\
		$\mu\nu$  & Belle~\cite{Abe:2007ws}    & $275\pm16\pm12$ & \\
		$\tau\nu$ ($\tau\to\pi\nu$) & 
			CLEO~\cite{Pedlar:2007za}   & $310\pm25\pm~8$ & $271\pm20\pm4$ \\
		$\tau\nu$ ($\tau\to e\nu\bar{\nu}$) & 
			CLEO~\cite{Ecklund:2007zm}  & $273\pm16\pm\;8$ & \\
		$\mu\nu$  & our average & $273\pm11$ & $271\pm10$ \\
		$\tau\nu$ & our average & $285\pm15$ & $272\pm13$ \\
		\hline\hline
	\end{tabular}
\end{table}
Treating both statistical and systematic uncertainties in quadrature, 
the average of the measurements in Table~\ref{tab:fDs} is
\begin{equation}
	\left.f_{D_s}\right|_{\rm expt~avg} = 272 \pm 8 \; \textrm{MeV}
\end{equation}
combining $\mu\nu$ and $\tau\nu$ and including new results reported by
CLEO at conferences through September 2008 \cite{CLEO:new}.
Separate averages for the two final states are in Table~\ref{tab:fDs}.

Now let us turn to lattice QCD calculations of $f_{D_s}$.
There are two calculations with 2+1 flavors of sea quarks, the first 
from the Fermilab Lattice and MILC Collaborations~\cite{Aubin:2005ar}
and more recently from the HPQCD Collaboration~\cite{Follana:2007uv}.
These are
\begin{equation}
	\left.f_{D_s}\right|_{\mathrm{HPQCD}} = 241 \pm  3 \; \textrm{MeV},
	\quad\quad
	\left.f_{D_s}\right|_{\mathrm{Fermilab\mbox{-}MILC}} = 
		249 \pm 11 \; \textrm{MeV},
	\label{eq:fDs-HPQCD} 
\end{equation}
where the Fermilab-MILC result is an update presented at this 
conference by Mackenzie~\cite{Mackenzie:2008ky}.
Both calculations use the improved staggered Asqtad action for 
\pagebreak
the sea quarks, taking advantage of the freely available MILC 
ensembles~\cite{Bernard:2001av}.
In the following, I use the average of the two results in
Eq.~(\ref{eq:fDs-HPQCD}) with no correlation, because the dominant
systematic errors differ.

The experimental average of $f_{D_s}$ lies 12.5\% above that of lattice 
QCD, and the significance of the discrepancy is
\begin{equation}
	3.5\sigma = 2.9\sigma \oplus 2.2\sigma,
	\label{eq:sigma}
\end{equation}
where the two entries on the right-hand side are for $\mu\nu$ and 
$\tau\nu$ separately.
Before the update from CLEO~\cite{CLEO:new} the discrepancy was
$3.8\sigma=2.7\sigma\oplus2.9\sigma$~\cite{Dobrescu:2008er}.
Reference~\cite{Rosner:2008yu} omits BaBar's result and reports
$3.4\sigma$, and with CLEO's update this approach yields $3.2\sigma$.
One should bear in mind that the yardstick for $\sigma$ is the
experimental \emph{statistical} error:
Were one to double HPQCD's error (without justification), the total
discrepancy would remain $3.0\sigma$.
Indeed, the same methods agree with experiment for $f_\pi$, $f_K$,
charmonium mass splittings, and especially $m_{D_s}$, $m_{D^+}$,
and~$f_{D^+}$ \cite{Follana:2007uv,Follana:2006rc}.

\section{New Physics}
\label{sec:np}

If the discrepancy cannot be traced to a fluctuation or error in either
the measurements or the calculation(s), then one should turn to
non-Standard physics as an explanation.
The new particles must be heavy to have escaped direct detection, 
so one may consider an effective Lagrangian
\begin{eqnarray}
	\mathcal{L}_{\mathrm{eff}} & = & 
	M^{-2}C_A^\ell(\bar{s}\gamma^\mu\gamma_5c)(\bar{\nu}_{\ell L}\gamma_\mu\ell_L) +
	M^{-2}C_P^\ell(\bar{s}\gamma_5c)(\bar{\nu}_{\ell L}\ell_R) -
	M^{-2}C_V^\ell(\bar{s}\gamma^\mu c)(\bar{\nu}_{\ell L}\gamma_\mu\ell_L)
	\nonumber \\ & + &
	M^{-2}C_S^\ell(\bar{s}c)(\bar{\nu}_{\ell L}\ell_R) +
	M^{-2}C_T^\ell(\bar{s}\sigma^{\mu\nu}c)(\bar{\nu}_{\ell L}\sigma_{\mu\nu}\ell_R)
	+ \textrm{H.c.},
	\label{eq:Leff}
\end{eqnarray}
where $M$ is a high mass scale.
This $\mathcal{L}_{\mathrm{eff}}$ extends the effective Lagrangian of
Ref.~\cite{Dobrescu:2008er} to include interactions that mediate 
$D\to K\ell\nu$.
The experiments do not identify the neutrino flavor or helicity, but
Eq.~(\ref{eq:Leff}) assumes $\bar{\nu}_{\ell L}$ to be of lepton 
flavor~$\ell$ and omits right-handed neutrinos.
In this way the resulting non-Standard amplitudes can interfere with the
Standard $W$-mediated amplitude and explain the sought-after effect of
10--15\% (in the amplitude).

These effective interactions change the rate for leptonic decay, by 
substituting in Eq.~(\ref{eq:Dslnu})
\begin{equation}
	 G_F V^*_{cs} m_\ell \to G_F V^*_{cs} m_\ell + 
		 \frac{C_A^\ell}{\sqrt{2}M^2} m_\ell + 
		 \frac{C_P^\ell}{\sqrt{2}M^2} \frac{m_{D_s}^2}{m_c+m_s}.
	\label{eq:lq-rate}
\end{equation}
Because (conventionally) $V_{cs}$ is real, one sees that one or both of 
$C_A^\ell$, $C_P^\ell$ must have a positive real part.
\pagebreak
If only one of these reduces the discrepancy to $1\sigma$, 
one can derive the bounds
\begin{equation}
	M(\Re C_A^\ell)^{-1/2} \lesssim  855~\textrm{GeV}, \quad\quad
	M(\Re C_P^\ell)^{-1/2} \lesssim 1070~\textrm{GeV} 
	\sqrt{m_\tau/m_\ell},
	\label{eq:estimation}
\end{equation}
updating Ref.~\cite{Dobrescu:2008er} to reflect CLEO's new preliminary
measurements and treating the $\mu\nu$ and $\tau\nu$ discrepancies as 
a single effect.

The effective Lagrangian can arise from the tree-level exchange of 
non-Standard particles, in which case $M$ is simply the new particle's 
mass.
Reference~\cite{Dobrescu:2008er} found a few possibilities.
One is the $s$-channel annihilation through a charged Higgs boson, in a new
model designed so that the Yukawa couplings satisfy
$y_s\ll y_c$ and $y_c,y_\tau\sim1$.
But this model also has $y_d<y_s$, thereby predicting a 10--15\%
deviation in the amplitude for $D^+\to\ell^+\nu$.
This is now disfavored, because CLEO's new measurement of
$f_{D^+}$~\cite{Eisenstein:2008sq} agrees perfectly with lattice
QCD~\cite{Aubin:2005ar,Follana:2007uv,Mackenzie:2008ky}.
Another candidate is the $t$-channel exchange of a charge $+\case{2}{3}$ 
leptoquark, which can arise in various ways, all of which are disfavored
by non-observation of $\tau\to\mu s\bar{s}$.
The most promising mechanism is the $u$-channel exchange of an
SU(2)-singlet, charge $-\third$ leptoquark, namely a particle with the
quantum numbers as a down-type scalar quark~$\tilde{d}$ in supersymmetric
models.
It couples via the $R$-violating Lagrangian
\begin{equation}
	\mathcal{L}_{\mathrm{LQ}} = \kappa_{2\ell} 
		\left(\bar{c}_L \ell_L^c - \bar{s}_L \nu_{\ell L}^c \right)\tilde{d} +
		\kappa^\prime_{2\ell} \, \bar{c}_R \ell_R^c\tilde{d} +
	\textrm{H.c.},
	\label{eq:lqL}
\end{equation}
where the superscript $c$ denotes charge conjugation,
and $\kappa_{2\ell}$ and $\kappa'_{2\ell}$ are complex parameters (in 
general, entries of Yukawa matrices).
When $M=m_{\tilde{d}}\gg m_{D_s}$ one can derive
$\mathcal{L}_{\mathrm{eff}}$ with
\begin{equation}
	C_A^\ell = C_V^\ell = \quarter |\kappa_{2\ell}|^2, \quad\quad
	C_P^\ell = C_S^\ell = \quarter \kappa_{2\ell}\kappa^{\prime *}_{2\ell}
		= -2C_T^\ell.
	\label{eq:CCC}
\end{equation}
If $\kappa_{2\ell}$ is independent of $\ell$ and either
$\kappa'_{2\ell}\propto m_\ell$ or
$|\kappa'_{2\ell}/\kappa_{2\ell}|\ll m_\ell m_c/m^2_{D_s}$, 
then these interactions could explain why the discrepancy appears
in both $\mu\nu$ and $\tau\nu$ channels.
Generalizations of Eq.~(\ref{eq:lqL}) appear in non-Standard models that 
modify the interference phase of $B^0_s$-$\bar{B}^0_s$~\cite{Kundu:2008ui}, 
explain quark masses~\cite{Dobrescu:2008sz},
induce deviations in $B^+_{(c)}\to\ell\nu$~\cite{Benbrik:2008ik},
generate neutrino masses~\cite{Dey:2008ht},
or enhance rare $D$ decays~\cite{Fajfer:2008tm}.

\section{Semileptonic Decays}
\label{sec:semileptonic}

To obtain further information about a possible non-Standard cause of the
effective $\bar{s}c\bar{\nu}\ell$ vertex, one can turn to other
processes.
One would be the production charmed quarks in neutrino scattering off
strange sea quarks in nucleons.
Another set consists of the semileptonic decays $D^0\to K^-\mu^+\nu_\mu$,
$D^+\to \bar{K}^0\mu^+\nu_\mu$, and their charge conjugates.
A full understanding of these decays will require lattice QCD 
calculations of the hadronic transition.

\begin{figure}[bp]
	\centering
	\includegraphics[width=\textwidth]{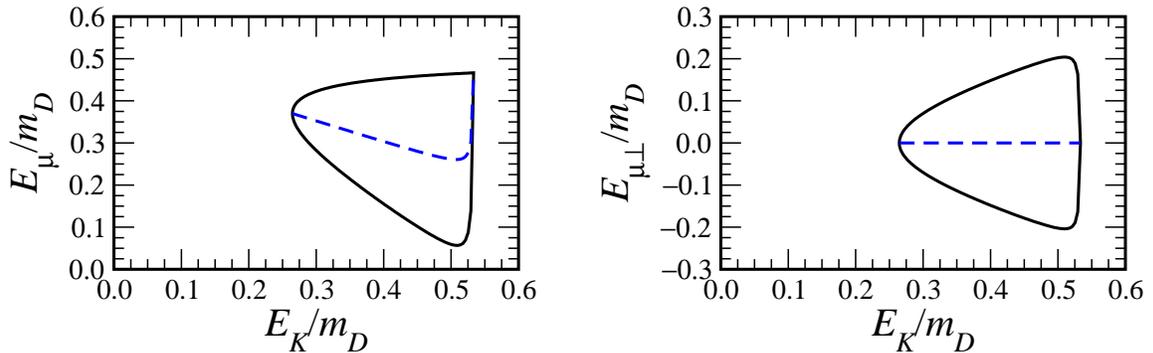}
	\caption[eq:Elperp]{Dalitz plots for $D\to K\ell\nu$, 
	with $m_\ell=m_\mu$.
	The left panel shows $E_\ell$, the lepton energy in the $D$ meson's
	rest frame; the right panel shows $E_{\ell\perp}$, defined in 
	Eq.~(\ref{eq:Elperp}).
	The dashed (blue) lines show $E_{\ell\perp}=0$.}
	\label{fig:dalitz}
\end{figure}

Let us start by reviewing the kinematics of three-body decays.
Let the $D$-meson, kaon, lepton, and neutrino 4-momenta be denoted
$p$, $k$, $\ell$, and $\nu$.
There are two Lorentz independent invariants, which may be taken to be
$E_\ell=p\cdot\ell/m_D$ and $E_K=p\cdot k/m_D$, namely the lepton and
kaon energies in the $D$ meson's rest frame.
Often instead of $E_K$ the mass-squared of the leptonic system is used,
$q^2=m_D^2+m_K^2-2m_DE_K$, $q=\ell+\nu=p-k$.
For brevity the formulae given below use both $E_K$ and $q^2$.
The kinematically allowed region is shown in the Dalitz plot,
Fig.~\ref{fig:dalitz}.
The discussion given below is somewhat simpler with the variable
\begin{equation}
	E_{\ell\perp} = \frac{p\cdot\ell}{m_D} - 
		\frac{p\cdot qq\cdot\ell}{m_Dq^2} = E_\ell - \half
		(m_D-E_K)\left(1+m_\ell^2/q^2\right),
	\label{eq:Elperp}
\end{equation}
and the allowed region, for fixed $E_K$ is
$-E_{\ell\perp}^{\rm max}\le E_{\ell\perp}\le E_{\ell\perp}^{\rm max}$,
$E_{\ell\perp}^{\rm max} = \half(1-m_\ell^2/q^2)\sqrt{E_K^2-m_K^2}$.
The allowed region for $E_K$ is
$ m_K\le E_K\le(m_D^2+m_K^2-m_\ell^2)/2m_D$,
or $m_\ell^2\le q^2\le(m_D-m_K)^2$.

The doubly-differential rate for $D\to K\ell\nu$ is
\begin{eqnarray}
	\frac{d^2\Gamma}{dE_K\,dE_{\ell\perp}} & = & \frac{m_D}{(2\pi)^3} 
		 \left\{ \vphantom{\frac{m_D^2-m_K^2}{q^2}}
		 \left[\left(E_K^2-m_K^2\right) \left(1-m_\ell^2/q^2\right) -
			4E_{\ell\perp}^2 \right]
		\left|G_FV_{cs}^* + G_V^\ell \right|^2 
		\left|f_+(q^2)\right|^2
	\right. \nonumber \\ & + & \left.
		\frac{q^2-m_\ell^2}{4m_D^2}
		\left|m_\ell\left(G_FV_{cs}^* + G_V^\ell\right) \frac{m_D^2-m_K^2}{q^2} + 
			G_S^\ell \frac{m_D^2-m_K^2}{m_c-m_s}\right|^2 
		\left|f_0(q^2)\right|^2
	\right. \nonumber \\ & + & \left.
		\left[\frac{m_\ell^2}{4m_D^2} \left(E_K^2-m_K^2\right) 
			\left(1-m_\ell^2/q^2\right) + 
		\frac{4q^2}{m_D^2}E_{\ell\perp}^2 \right]
		\left|G_T^\ell\right|^2 \left|f_2(q^2)\right|^2
	\right. \label{eq:dGammadEKdEl} \\ & - & \left.
		\frac{2m_\ell}{m_D} \left(E_K^2-m_K^2\right) \left(1-m_\ell^2/q^2\right)
		\Re\left[\left(G_FV_{cs}^* + G_V^\ell\right)
		G_T^{\ell*} f_+(q^2) f_2^*(q^2)\right] 
	\right. \nonumber \\ & - & \left.
		\frac{2m_\ell}{m_D} E_{\ell\perp} \Re\left[\left(
		m_\ell \left(G_FV_{cs}^* + G_V^\ell\right)
		\frac{m_D^2-m_K^2}{q^2} + 
		G_S^\ell \frac{m_D^2-m_K^2}{m_c-m_s} \right)
	\times \right.\right. \nonumber \\ & & \hspace*{6em} \left.\left.
		\left(G_FV_{cs} + G_V^{\ell*}\right) 
		f_0(q^2) f_+^*(q^2)\vphantom{\frac{m_D^2-m_K^2}{q^2}}\right]
	\right. \nonumber \\ & + & \left.
		\frac{2q^2}{m_D^2} E_{\ell\perp} \Re\left[\left(
		m_\ell \left(G_FV_{cs}^* + G_V^\ell\right)
			\frac{m_D^2-m_K^2}{q^2} + 
		G_S^\ell \frac{m_D^2-m_K^2}{m_c-m_s} \right)
		G_T^{\ell*} f_0(q^2) f_2^*(q^2) 
		\vphantom{\frac{m_D^2-m_K^2}{q^2}}\right]
	\right\}, \nonumber
\end{eqnarray}
where $G_{V,S,T}^\ell=C_{V,S,T}^\ell/\sqrt{2}M^2$, and
the form factors $f_+$, $f_0$, and $f_2$ are defined via
\begin{eqnarray}
	\langle K(k)|\bar{s}\gamma^\mu c|D(p)\rangle & = &
		\left(p^\mu+k^\mu-\frac{m_D^2-m_K^2}{q^2}q^\mu\right) f_+(q^2) +
		\frac{m_D^2-m_K^2}{q^2}q^\mu f_0(q^2),
	\label{eq:f+} \\
	\langle K(k)|\bar{s}\sigma^{\mu\nu} c|D(p)\rangle & = & i m_D^{-1}
		(p^\mu k^\nu-p^\nu k^\mu) f_2(q^2), 
	\label{eq:f2} \\
	\langle K(k)|\bar{s} c|D(p)\rangle & = &
		\frac{m_D^2-m_K^2}{m_c-m_s} f_0(q^2),
	\label{eq:f0}
\end{eqnarray}
and $f_0$ appears for both the vector and scalar currents owing to~CVC.
Integrating over lepton energy
\begin{eqnarray}
	\frac{d\Gamma}{dE_K} & = & \frac{m_D}{(2\pi)^3}
		\sqrt{E_K^2-m_K^2} \left(1-\frac{m_\ell^2}{q^2} \right)^2 
		\left\{ (E_K^2-m_K^2) \frac{2q^2 + m_\ell^2}{3q^2}
		\left|G_FV_{cs}^* + G_V^\ell \right|^2 \left|f_+(q^2)\right|^2
	\right. \nonumber \\ & + & \left.
		\frac{q^2}{4m_D^2}
		\left|m_\ell \left(G_FV_{cs}^* + G_V^\ell\right)
			\frac{m_D^2-m_K^2}{q^2} + 
		G_S^\ell \frac{m_D^2-m_K^2}{m_c-m_s}\right|^2
		\left|f_0(q^2)\right|^2
	\right. \label{eq:dGammadEK} \\ & + & \left.
		(E_K^2-m_K^2) \frac{q^2+2m_\ell^2}{3m_D^2}
		\left|G_T^\ell\right|^2 \left|f_2(q^2)\right|^2
	\right. \nonumber \\ & - & \left.
		2\frac{m_\ell}{m_D} (E_K^2-m_K^2)
		\Re\left[\left(G_FV_{cs}^* + G_V^\ell\right)
		G_T^{\ell*} f_+(q^2) f_2^*(q^2)\right] 
	\right\}. \nonumber
\end{eqnarray}
Note that two terms with interference between form factors vanish after
integration: the variable~$E_{\ell\perp}$ renders this feature
especially transparent.
We shall use these formulae to diagnose how new interactions mediating
$D_s\to\ell\nu$ would alter the semileptonic rate and differential
distributions.

In many non-Standard models, including those with the charge $-\third$ 
leptoquark~$\tilde{d}$, one finds $C_V^\ell=C_A^\ell$ and 
$C_S^\ell=C_P^\ell$ (cf.\ Eq.~(\ref{eq:CCC})).
If the $f_{D_s}$ puzzle is solved by the $C_A^\ell$ interaction, then 
the $C_V^\ell$ interaction generates a similarly large enhancement in the 
semileptonic rate.
This consequence is easily seen from the $|f_+|^2$ contribution to the 
\pagebreak
rate: the other Standard contribution, with $|f_0|^2$, is suppressed by 
$(m_\ell/m_D)^2$, which is $3\times10^{-3}$ ($7\times10^{-8}$) for 
$\mu$~($e$).
On the other hand, if the $f_{D_s}$ puzzle is solved by the $C_P^\ell$ 
interaction, then it will be difficult to observe the companion  
semileptonic contribution.
To enhance both $D_s\to\tau\nu$ and $D_s\to\mu\nu$, some mechanism
should lead to $C_P^\ell\propto m_\ell$, and then (in the leptoquark
example) $C_S^\ell\propto m_\ell$,  $C_T^\ell\propto m_\ell$ also.
In that case, the non-Standard contributions are small corrections to a 
suppressed contribution.

The doubly-differential rate suggests a challenging way to observe the
effects of non-vanishing $C_S^\ell$ and $C_T^\ell$.
In the asymmetry
\begin{equation}
	\mathcal{A}_\perp = \frac{\Gamma(E_{\ell\perp}>0)-\Gamma(E_{\ell\perp}<0)}%
{\Gamma(E_{\ell\perp}>0)+\Gamma(E_{\ell\perp}<0)} = 
\frac{N(E_{\ell\perp}>0)-N(E_{\ell\perp}<0)}%
{N(E_{\ell\perp}>0)+N(E_{\ell\perp}<0)}
\end{equation}
everything but the last two lines of Eq.~(\ref{eq:dGammadEKdEl}) cancels.
To obtain a 7\% measurement of $\mathcal{A}_\perp$, one would need
around $10^7$ semimuonic events in each half of the modified Dalitz
plane.
One can generalize this asymmetry to any region in the 
$E_K$-$E_{\ell\perp}$ plane that is symmetric about $E_{\ell\perp}=0$,
or to any moment of the distribution odd in~$E_{\ell\perp}$.
For example, when the kaon momentum is low, phase space naturally
suppresses the $|f_+|^2$ contribution, perhaps helpfully.

\section{Conclusions}
\label{sec:end}

From Eqs.~(\ref{eq:dGammadEKdEl}) and~(\ref{eq:dGammadEK}) one sees that
the first concern of future lattice calculations is to improve on the
7\% uncertainty of the only 2+1 flavor calculation of
$f_+(q^2)$~\cite{Aubin:2004ej}.
With current semielectronic 
measurements~\cite{Aubert:2007wg,Cronin-Hennessy:2007se}, one could 
test for new contributions to the $\bar{s}c\bar{\nu}_ee$ vertex (for 
which there is not yet any evidence).
Semimuonic measurements are needed for a direct test of the $f_{D_s}$ 
puzzle.
Once the event yields become high enough to measure $\mathcal{A}_\perp$,
it will be necessary to have accurate calculations of the scalar and
tensor form factors, $f_0$ and~$f_2$.
CLEO and the $B$ factories have somewhat more data to analyze, and
BES-III should record thousands of events~\cite{Li:2008wv}.


\acknowledgments
I would like to thank Bogdan Dobrescu for an enjoyable
collaboration, as well as
Dan Cronin-Hennessy,
Christine Davies,
Eduardo Follana,
Paddy Fox,
Jack Laiho,
Peter Lepage,
Elliot Lipeles,
Enrico Lunghi,
Paul Mackenzie,
Sheldon Stone,
and 
Ruth Van de Water for useful conversations.

\appendix

\section{All Semileptonic Formulas}

To treat the SM and NP efficiently, we shall now write the effective 
Lagrangian as
\begin{eqnarray}
	\mathcal{L}_{\mathrm{eff}} & = & 
	M^{-2}\bar{C}_A^\ell(\bar{s}\gamma^\mu\gamma_5c)(\bar{\nu}_L\gamma_\mu\ell_L) -
	M^{-2}\bar{C}_V^\ell(\bar{s}\gamma^\mu c)(\bar{\nu}_L\gamma_\mu\ell_L) +
	M^{-2}C_P^\ell(\bar{s}\gamma_5c)(\bar{\nu}_L\ell_R)
	\nonumber \\ & + &
	M^{-2}C_S^\ell(\bar{s}c)(\bar{\nu}_L\ell_R) +
	M^{-2}C_T^\ell(\bar{s}\sigma^{\mu\nu}c)(\bar{\nu}_L\sigma_{\mu\nu}\ell_R)
	+ \textrm{H.c.},
	\label{eq:barLeff}
\end{eqnarray}
where
\begin{equation}
	\bar{C}_{V,A}^\ell = \sqrt{2}M^2G_FV_{cs}^* + C_{V,A}^\ell,\quad\quad
	\frac{\bar{C}_{V,A}^\ell}{\sqrt{2}M^2} = 
		G_FV_{cs}^* + \frac{C_{V,A}^\ell}{\sqrt{2}M^2}.
\end{equation}
This $\mathcal{L}_{\mathrm{eff}}$ mediates $D\to K\ell\nu$.
(By analogy, one can extend this to the semileptonic decay of any 
pseudoscalar meson.)
Let the $D$-meson, kaon, lepton, and neutrino 4-momenta be denoted
$p$, $k$, $\ell$, and $\nu$, as above.
The amplitude is
\begin{eqnarray}
	\langle\ell\nu K|i\mathcal{L}_{\mathrm{eff}}|D\rangle = iM^{-2} 
		\bar{u}(\nu)\half(1+\gamma_5) \left[ 
			C_S^\ell \langle K|\bar{s} c|D\rangle \right. & - &
			\bar{C}_V^\ell \gamma_\mu 
				\langle K|\bar{s}\gamma^\mu c|D\rangle 
		\label{eq:amplitude} \\ & + & \left.
			C_T^\ell \sigma_{\mu\nu}
				\langle K|\bar{s}\sigma^{\mu\nu} c|D\rangle 
		\right] v(\ell). \nonumber
\end{eqnarray}
The hadronic matrix elements are re-expressed as form factors in
Eqs.~(\ref{eq:f+})--(\ref{eq:f2}).
For the leptons we require the spinor combinations
\begin{eqnarray}
	\bar{u}(\nu)\half(1+\gamma_5)(\slsh{p}+\slsh{k})_\perp v(\ell) & = &
		2\bar{u}(\nu)\slsh{p}_\perp\half(1-\gamma_5)v(\ell), 
	\label{eq:spinorV} \\
	\bar{u}(\nu)\half(1+\gamma_5)\slsh{q}v(\ell) & = & -
		m_\ell\bar{u}(\nu)\half(1+\gamma_5)v(\ell), \\
	(p^\mu k^\nu - p^\nu k^\mu)
		\bar{u}(\nu)\half(1+\gamma_5)i\sigma_{\mu\nu}v(\ell) & = & 
		-2\bar{u}(\nu)\slsh{p}_\perp\slsh{q}\half(1+\gamma_5)v(\ell),
	\label{eq:spinorT}
\end{eqnarray}
where for any four-vector $r$, $r_\perp^\mu=r^\mu-(r\cdot q/q^2)q^\mu$.
Inserting Eqs.~(\ref{eq:f+})--(\ref{eq:f2})
and~(\ref{eq:spinorV})--(\ref{eq:spinorT}) into
Eq.~(\ref{eq:amplitude}),
\begin{eqnarray}
	\langle\ell\nu K|i\mathcal{L}_{\mathrm{eff}}|D\rangle & = & iM^{-2} 
		\left\{ \left( m_\ell \bar{C}_V^\ell \frac{m_D^2-m_K^2}{q^2} + 
			C_S^\ell\frac{m_D^2-m_K^2}{m_c-m_s} \right)
		\bar{u}(\nu)\half(1+\gamma_5)v(\ell) f_0(q^2)
		\right. \\ & - & \left. 
			2\bar{C}_V^\ell 
			\bar{u}(\nu)\slsh{p}_\perp\half(1-\gamma_5)v(\ell) f_+(q^2) - 
			2C_T^\ell m_D^{-1}
			\bar{u}(\nu)\slsh{p}_\perp\slsh{q}\half(1+\gamma_5)v(\ell)
			f_2(q^2) \vphantom{\frac{m_D^2}{q^2}} \right\}. \nonumber
\end{eqnarray}

The differential rate is (see PDG)
\begin{equation}
	\frac{d^2\Gamma}{dE_K\,dE_\ell} = 
		\frac{1}{(2\pi)^3} \frac{1}{8m_D} \sum_{\rm spins}
		|\langle\ell\nu K|i\mathcal{L}_{\mathrm{eff}}|D\rangle|^2,
\end{equation}
where $E_K=p\cdot k/m_D$ and $E_\ell=p\cdot\ell/m_D$ are the energies 
of the kaon and lepton in the rest frame of the $D$ meson.
To sum over lepton and neutrino polarization states, we need
\begin{eqnarray}
	\sum_{\mathrm{spins}} \bar{u}(\nu)\half(1+\gamma_5)v(\ell)
		\bar{v}(\ell)\half(1-\gamma_5)u(\nu) & = & 2\nu\cdot\ell =
		q^2-m_\ell^2, \\
	\sum_{\mathrm{spins}} \bar{u}(\nu)\slsh{p}_\perp\half(1-\gamma_5)v(\ell)
		\bar{v}(\ell)\half(1+\gamma_5)\slsh{p}_\perp u(\nu) & = &
		-p_\perp^2(q^2-m_\ell^2)-(2p_\perp\cdot\ell)^2, \\
	\sum_{\mathrm{spins}}
		\bar{u}(\nu)\slsh{p}_\perp\slsh{q}\half(1+\gamma_5)v(\ell)
		\bar{v}(\ell)\half(1-\gamma_5)\slsh{q}\slsh{p}_\perp u(\nu) & = &
		q^2 (2p_\perp\cdot\ell)^2 - m_\ell^2 p_\perp^2 (q^2-m_\ell^2), \\
	\sum_{\mathrm{spins}} \bar{u}(\nu)\half(1+\gamma_5)v(\ell)
		\bar{v}(\ell)\half(1+\gamma_5)\slsh{p}_\perp u(\nu) & = & 
		2 m_\ell p_\perp\cdot\ell, \\	
	\sum_{\mathrm{spins}} \bar{u}(\nu)\half(1+\gamma_5)v(\ell)
		\bar{v}(\ell)\half(1-\gamma_5)\slsh{q}\slsh{p}_\perp u(\nu) & = & 
		-2q^2 p_\perp\cdot\ell, \\
	\sum_{\mathrm{spins}} \bar{u}(\nu)\slsh{p}_\perp\half(1-\gamma_5)v(\ell)
		\bar{v}(\ell)\half(1-\gamma_5)\slsh{q}\slsh{p}_\perp u(\nu) & = & 
		m_\ell p_\perp^2 (q^2-m_\ell^2), 
\end{eqnarray}
in which $p_\perp^2=-(E_K^2-m_K^2)m_D^2/q^2$.
Note that $E_K^2-m_K^2$ is nothing but the kaon's squared 
three-momentum in the $D$-meson rest frame.
It is convenient to change from the variable
$E_\ell$ to $E_{\ell\perp}=p_\perp\cdot\ell/m_D$,
defined in the text.
With this variable, it is easy to see which contributions vanish
after integrating over lepton energy.

The doubly-differential rate is then
\begin{eqnarray}
	\frac{d^2\Gamma}{dE_K\,dE_{\ell\perp}} & = & \frac{1}{(2\pi)^3} 
		\frac{m_D}{2M^4} \left\{ 
		 \left[(E_K^2-m_K^2)\frac{q^2-m_\ell^2}{q^2} -
			4E_{\ell\perp}^2 \right]
		\left|\bar{C}_V^\ell\right|^2 |f_+(q^2)|^2
	\right. \nonumber \\ & + & \left.
		\frac{q^2-m_\ell^2}{4m_D^2}
		\left|m_\ell \bar{C}_V^\ell\frac{m_D^2-m_K^2}{q^2} + 
		C_S^\ell\frac{m_D^2-m_K^2}{m_c-m_s}\right|^2
		|f_0(q^2)|^2
	\right. \nonumber \\ & + & \left.
		\left[\frac{m_\ell^2}{m_D^2} (E_K^2-m_K^2) \frac{q^2-m_\ell^2}{q^2} + 
		\frac{4q^2}{m_D^2}E_{\ell\perp}^2 \right]
		\left|C_T^\ell\right|^2 |f_2(q^2)|^2
	\right. \\ & - & \left.
		\frac{2m_\ell}{m_D} (E_K^2-m_K^2) \frac{q^2-m_\ell^2}{q^2}
		\Re\left[\bar{C}_V^\ell
		{C_T^\ell}^* f_+(q^2) f_2^*(q^2)\right] 
	\right. \nonumber \\ & - & \left.
		\frac{2m_\ell}{m_D} E_{\ell\perp} \Re\left[\left(
		m_\ell \bar{C}_V^\ell\frac{m_D^2-m_K^2}{q^2} + 
		C_S^\ell\frac{m_D^2-m_K^2}{m_c-m_s} \right)
		\bar{C}_V^{\ell*} f_0(q^2) f_+^*(q^2)\right]
	\right. \nonumber \\ & + & \left.
		\frac{2q^2}{m_D^2} E_{\ell\perp} \Re\left[\left(
		m_\ell \bar{C}_V^\ell\frac{m_D^2-m_K^2}{q^2} + 
		C_S^\ell\frac{m_D^2-m_K^2}{m_c-m_s} \right)
		{C_T^\ell}^* f_0(q^2) f_2^*(q^2)\right]
	\right\}. \nonumber
\end{eqnarray}
The singly-differential rate is 
\begin{eqnarray}
	\frac{d\Gamma}{dE_K} & = & \frac{1}{(2\pi)^3} \frac{m_D}{2M^4} 
		\sqrt{E_K^2-m_K^2} \left(1-\frac{m_\ell^2}{q^2} \right)^2 
		\left\{ (E_K^2-m_K^2) \frac{2q^2 + m_\ell^2}{3q^2}
			\left|\bar{C}_V^\ell\right|^2 |f_+(q^2)|^2
	\right. \nonumber \\ & + & \left.
		\frac{q^2}{4m_D^2}
		\left|m_\ell \bar{C}_V^\ell\frac{m_D^2-m_K^2}{q^2} + 
		C_S^\ell\frac{m_D^2-m_K^2}{m_c-m_s}\right|^2
		|f_0(q^2)|^2
	\right. \\ & + & \left.
		(E_K^2-m_K^2) \frac{q^2+2m_\ell^2}{3m_D^2}
		\left|C_T^\ell\right|^2 |f_2(q^2)|^2 -
		\frac{2m_\ell}{m_D} (E_K^2-m_K^2)
		\Re\left[\bar{C}_V^\ell
		{C_T^\ell}^* f_+(q^2) f_2^*(q^2)\right] 
	\right\} \nonumber \\
	 & = & \frac{m_D}{(2\pi)^3}
		\sqrt{E_K^2-m_K^2} \left(1-\frac{m_\ell^2}{q^2} \right)^2 
		\left\{ (E_K^2-m_K^2) \frac{2q^2 + m_\ell^2}{3q^2}
		\left|G_FV_{cs}^* + \frac{C_V^\ell}{\sqrt{2}M^2}\right|^2 |f_+(q^2)|^2
	\right. \nonumber \\ & + & \left.
		\frac{q^2}{4m_D^2}
		\left|m_\ell \left(G_FV_{cs}^* + \frac{C_V^\ell}{\sqrt{2}M^2}\right)
			\frac{m_D^2-m_K^2}{q^2} + 
		\frac{C_S^\ell}{\sqrt{2}M^2}\frac{m_D^2-m_K^2}{m_c-m_s}\right|^2
		|f_0(q^2)|^2
	\right. \nonumber \\ & + & \left.
		(E_K^2-m_K^2) \frac{q^2+2m_\ell^2}{3m_D^2}
		\left|\frac{C_T^\ell}{\sqrt{2}M^2}\right|^2 |f_2(q^2)|^2
	\right. \\ & - & \left.
		2\frac{m_\ell}{m_D} (E_K^2-m_K^2)
		\Re\left[\left(G_FV_{cs}^* + \frac{C_V^\ell}{\sqrt{2}M^2}\right)
		\frac{{C_T^\ell}^*}{\sqrt{2}M^2} f_+(q^2) f_2^*(q^2)\right] 
	\right\}. \nonumber
\end{eqnarray}

\end{document}